\documentclass{aa}

\def\aap{A\&A}
\def\aas{A\&AS}
\def\apj{ApJ}
\def\aps{ApJS}
\def\apss{Ap\&SS}
\def\asr{Adv.~Space Res.}
\def\ast{J.~Atm.\ Solar Terr.\ Phys.}
\def\bac{Bull.~Astron.~Inst.~Czechosl.}
\def\grl{Geophys.\ Res.\ Lett.}
\def\ica{Icarus}

\def\jgr{J.~Geophys.~Res.}
\def\jpc{J.~Phys.\ Chem.\ Ref.\ Data}
\def\mon{Moon}
\def\nat{Nature}
\def\prv{Physical Review}
\def\pss{Planet.~Space~Sci.}
\def\<#1>{\relax}
\usepackage{natbib}
\bibpunct{(}{)}{;}{a}{}{,}
\usepackage{graphicx}
\usepackage{psfig}

\begin{document}

\title{Discovery of X--rays from Mars with Chandra}

\author{K. Dennerl}

\institute{Max--Planck--Institut f\"ur extraterrestrische Physik,
           Giessenbachstra{\ss}e, 85748 Garching, Germany}

\offprints{K. Dennerl, \email{kod@mpe.mpg.de}}

\date{Received 16 July 2002 / Accepted 31 July 2002}

\abstract{On 4 July 2001, X--rays from Mars were detected for the first time.
The observation was performed with the ACIS--I detector onboard
Chandra and yielded data of high spatial and temporal resolution, together
with spectral information. Mars is clearly detected as an almost fully
illuminated disk, with an indication of limb brightening at the sunward
side, accompanied by some fading on the opposite side. The morphology and the
X--ray luminosity of $\sim4\mbox{ MW}$ are fully consistent with fluorescent
scattering of solar X--rays in the upper Mars atmosphere. The X--ray spectrum
is dominated by a single narrow emission line, which is most likely caused by
O--K$_{\alpha}$ fluorescence. No evidence for temporal variability is found.
This is in agreement with the solar X--ray flux, which was almost constant
during the observation. In addition to the X--ray fluorescence, there is
evidence for an additional source of X--ray emission, indicated by a faint
X--ray halo which can be traced to about three Mars radii, and by an
additional component in the X--ray spectrum of Mars, which has a similar
spectral shape as the halo. Within the available limited statistics, the
spectrum of this component can be characterized by 0.2~keV thermal
bremsstrahlung emission. This is indicative of charge exchange interactions
between highly charged heavy ions in the solar wind and exospheric hydrogen
and oxygen around Mars. Although the observation was performed at the onset of
a global dust storm, no evidence for dust--related X--ray emission was found.
}

\authorrunning{K. Dennerl}

\maketitle

\begin{figure}
\centerline{\resizebox{1.0\hsize}{!}{\psfig{file=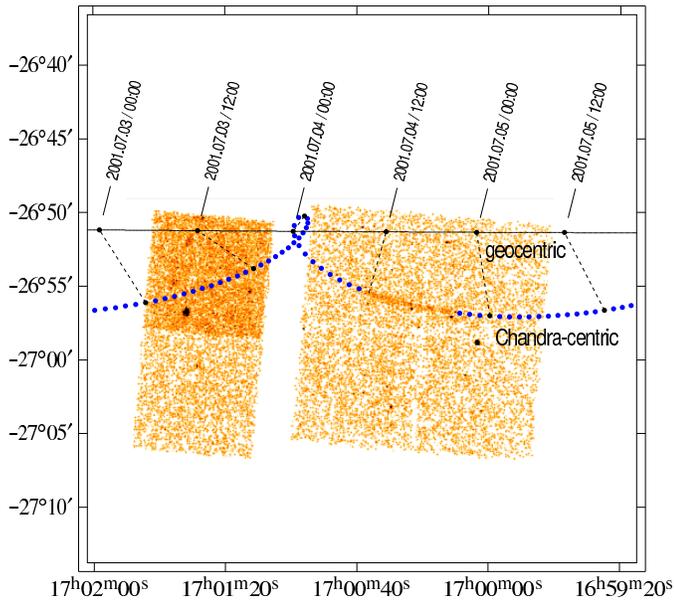,clip=}}}
\caption{Apparent motion of Mars in celestial coordinates (J\,2000), as seen
from Earth and Chandra. The solid line shows the geocentric path of Mars, with
its position indicated every 12 hours. Dashed lines illustrate the parallactic
shift at these times due to the Chandra orbit. Chandra--centric positions are
shown every hour by dots, which match the apparent size of Mars. The loop at 4
July, 0~UT reflects the perigee passage of Chandra. Superimposed on this
diagram is the Chandra image in celestial coordinates, obtained at
$E<1.0\mbox{ keV}$ from all events with standard grades. The sequence of dots
in the Chandra--centric curve is interrupted to show the curved soft X--ray
trail of Mars on the ACIS--I array. In addition to Mars, several point sources
show up in the Chandra image. The photons from these sources were excluded
in the subsequent analysis.
}
\label{marspath}
\end{figure}

\section{Introduction}

During the last years our knowledge about the X--ray properties of solar
system objects was considerably enhanced. While the Sun was long known to be
an X--ray source \citep{51phr008}\<Friedman..>, X--rays from the Earth
\citep{68jgr012}\<Grader>, the Moon
\citep{74mon001,91nat046}\<Gorenstein..>\<Schmitt..>, and Jupiter
\citep{83jgr013}\<Metzger..> were detected a few decades later. In 1996 comets
were discovered as a new class of X--ray sources
\citep[e.g.][]{96sci001,97sci002,97apj353}\<Lisse.. Dennerl.. Mumma..>.
Prompted by this discovery, \cite{00apj363}\<Cravens> found evidence for
X--ray emission from the heliosphere. A marginal X--ray detection of Saturn
was reported by \cite{00aap289}\<Ness..>, and, more recently, X--ray emission
was discovered from the Galilean satellites Io and Europa and the Io plasma
torus \citep{02apj367}\<Elsner>, and from Venus \citep{02aap288}\<Dennerl..>.
Here the first detection of X--ray emission from Mars is reported.

\begin{table*}
\caption[]{Journal of observations and observing geometry}
\label{obscxo}
\begin{tabular}{ccccccccccc}
\hline
\noalign{\smallskip}
obsid & date & time & exp time & instrument & $r$ & $\Delta$ & phase &
elong & diam \\
 & 2001 & [UT] & [s] & & [AU] & [AU] & [$^{\circ}$] & [$^{\circ}$] & [$''$] \\
\hline
\noalign{\medskip}
1861 & July 4 & 11:47:39\,--\,21:00:30 & 33\,171 & ACIS--I & 1.446 & 0.462 &
                                         18.2 & 153.7 & 20.3 \\
\noalign{\smallskip}
\hline
\end{tabular} \\[2ex]
obsid: Chandra observation identifier, exp time: exposure time,
$r$: distance from Sun,\\
$\Delta$: distance from Earth,
phase: angle Sun--Mars--Earth, elong: angle Sun--Earth--Mars,\\
diam: apparent diameter
%
%
\end{table*}

\section{Observation and data analysis}

Mars was observed on 4 July 2001, during the first opposition after the launch
of Chandra. Details about the observing geometry can be found in
Table~\ref{obscxo}. There were already expectations that Mars would be an
X--ray source, though a very faint one (Sect.\,4). Thus the prime goal for
this observation was to get an unambiguous detection. Direct imaging onto the
CCDs of the ACIS--I array, operated in very faint mode, was chosen as the
observing mode, in order to get spectral information, which can also be used
for suppressing the background efficiently. Due to the high surface brightness
of Mars ($4.1\mbox{ mag arcsec}^{-2}$), direct imaging with the more
sensitive back--illuminated ACIS--S3 CCD might have led to problems with
contamination by optical light.

\smallskip
The observing technique is illustrated in Figure~\ref{marspath}. Chandra was
pointed such that Mars would be close to the nominal aimpoint in I3 during the
middle of the observation, to get the sharpest possible image and a minimum of
vignetting. As the CCDs were read out every 3.2~s, there was no need for
continuous tracking. The photons were individually transformed into the rest
frame of Mars, using the geocentric ephemeris of Mars, computed with the JPL
ephemeris calculator.%
\footnote{available at \tt http://ssd.jpl.nasa.gov/cgi-bin/eph}
Correction for the parallax of Chandra was done with the orbit ephemeris of the
delivered data set. For the whole analysis events with Chandra standard grades
were used. The ACIS particle background was reduced by screening out events
with significant flux in border pixels of the $5\times5$ event islands.

\smallskip
In order to avoid contamination of the X--ray signal by unrelated point
sources, photons within the point spread function of such sources
were removed. Due to the high spatial resolution of the Chandra X--ray
telescope, this can be done very efficiently. The insensitive areas, which
are created by this method in the celestial reference frame, are very small
(cf.\,Fig.\,\ref{marspath}). After transformation into the rest frame of Mars,
they become diluted along the proper motion direction of Mars, causing an
almost negligible reduction of the effective exposure along such streaks.
In order to avoid inhomogeneous sensitivity caused by the gaps between the
CCDs along the path of Mars (Fig.\,\ref{marspath}), the analysis was
restricted to photons within $100''$ from the center of Mars.

\begin{figure}
\centerline{\resizebox{1.0\hsize}{!}{\psfig{file=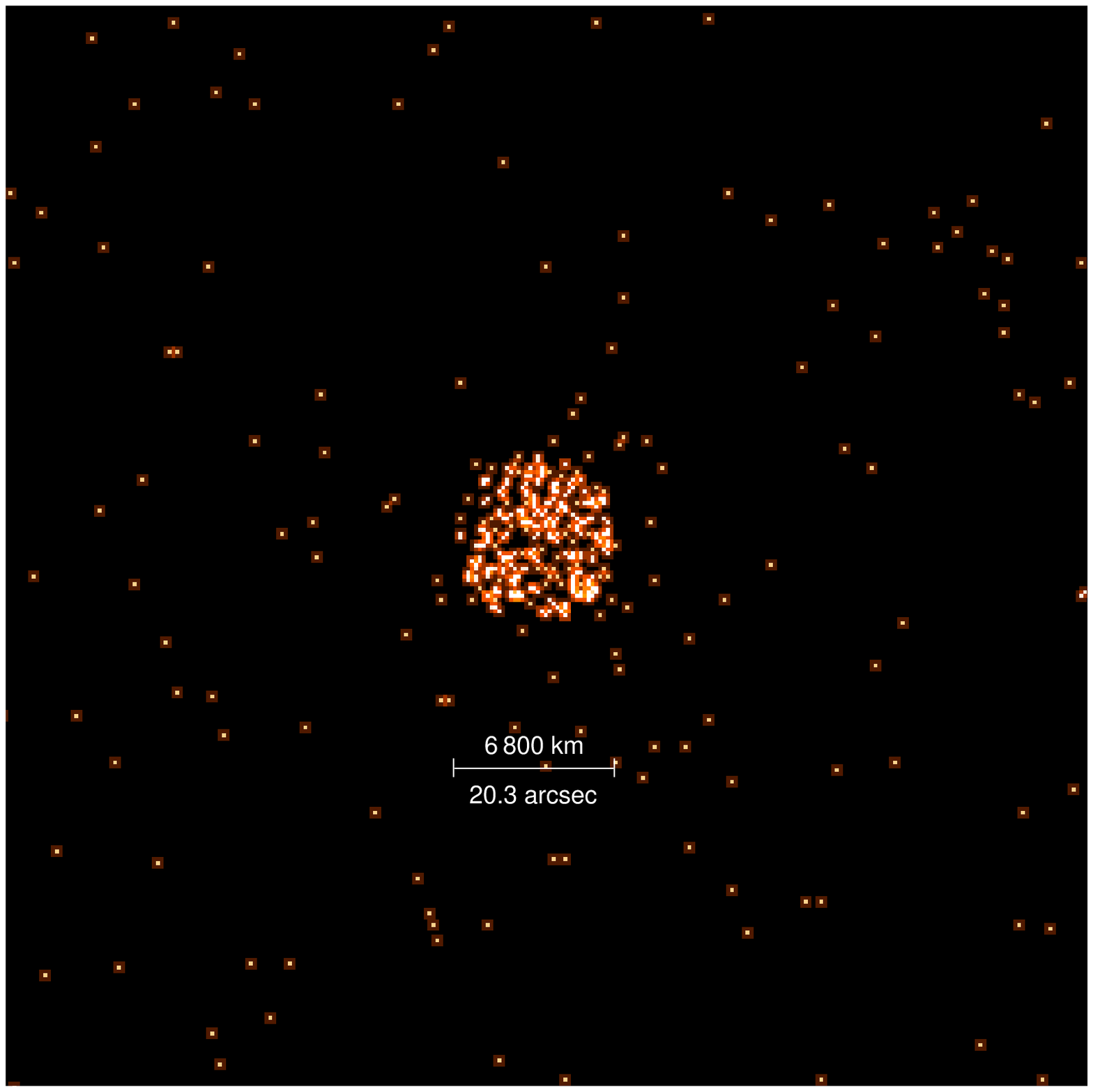,clip=}}}
\caption{First X--ray image of Mars, obtained with Chandra
\mbox{ACIS--I} on 4 July 2001.
Only photons in the instrumental energy range $E=0.40$\,--\,0.73~keV
were selected and transformed into the rest frame of Mars. Trails of
point sources were removed.
}
\label{mars2}
\end{figure}

\begin{figure}
\hbox to \hsize{\vbox{\hsize=1.0\hsize
\resizebox{0.9\hsize}{!}{\psfig{file=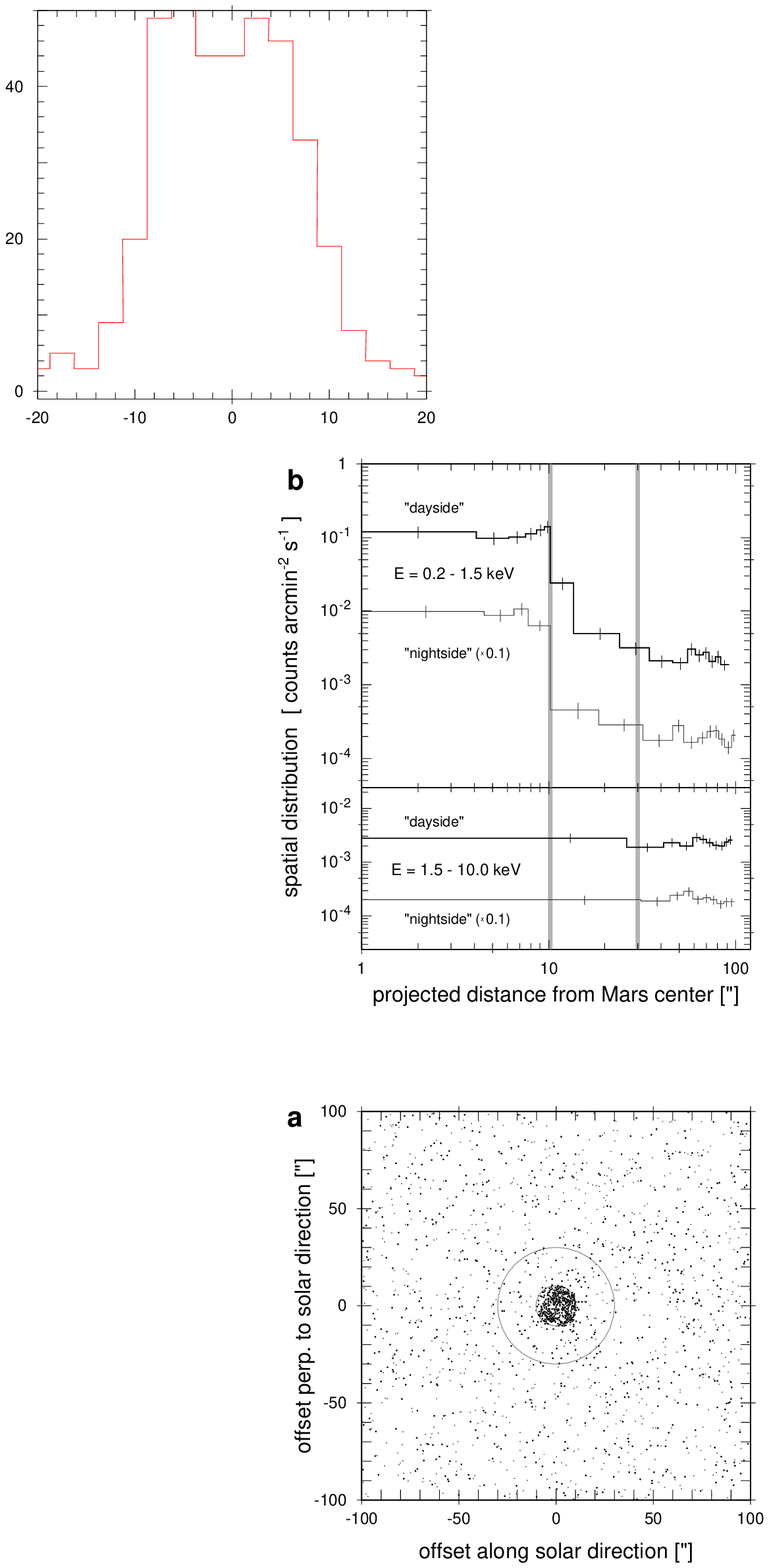,%
bbllx=295pt,bblly=72pt,bburx=515pt,bbury=285pt,clip=}} \\
\resizebox{0.9\hsize}{!}{\psfig{file=f03.eps,%
bbllx=295pt,bblly=315pt,bburx=515pt,bbury=566pt,clip=}}
}\hfil}
\caption{Spatial distribution of photons around Mars in the ACIS--I
observation, after removing the trails of point sources.
{\bf a)} All photons in the energy range
0.2\,--\,10.0~keV; those with $E\le1.5\mbox{ keV}$
are marked with large dots, while photons with $E>1.5\mbox{ keV}$ are
plotted as small dots. In some cases the dots have been
slightly shifted (by typically less than $1\arcsec$) to minimize overlaps.
The inner circle, with $r=10\farcs2$, marks the geometric size of Mars, while
the outer one, at $r=30''$, illustrates the extent of the soft X--ray halo.
{\bf b)} Spatial distribution of the photons in the soft
($E=0.2$\,--\,1.5~keV) and hard ($E=1.5$\,--\,10.0~keV) energy range,
in terms of surface brightness along radial rings around Mars,
separately for the ``dayside" (offset along projected solar direction $>0$)
and the ``nightside" (offset $<0$); note, however, that the phase angle
was only $18\fdg2$. For better clarity the nightside histograms were shifted
by one decade downward. The bin size was adaptively determined so that each
bin contains at least 28 counts. The thick vertical lines mark the radii
$10\farcs2$ and $30''$ of the circles in (a).
}
\label{ms4spa}
\end{figure}

\section{Observational results}

\subsection{Morphology}

Mars is clearly detected in the Chandra image (Fig.\,\ref{mars2}), and,
although the photon statistics are low, some general information about the
brightness distribution across the disk can be derived. Figure~\ref{ms4spa}b
shows the average surface brightness as a function of the distance from the
center of Mars. In the (instrumental) energy range 0.2\,--\,1.5~keV, a limb
brightening by $\sim25\%$ is indicated on the sunward limb
(Fig.\,\ref{ms4spa}b,``dayside''), while a darkening is seen at the opposite
limb (``nightside'').

\smallskip
At larger distances, the radial brightness profiles show some evidence for a
soft X--ray halo around Mars, extending out to $\sim30''$ (3 Mars radii), both
on the ``dayside'' and the ``nightside''. No evidence for such a halo is seen
above 1.5~keV. The halo is most pronounced in the energy range
0.5\,--\,1.0~keV, where the annulus between $r=11''$ and $r=30''$ around Mars
contains $34.6\pm8.4$ excess counts relative to the background expected for
this area, as determined from the $r=50''$ to $r=100''$ annulus around Mars.

\subsection{Spectrum}

\begin{figure}
\centerline{\resizebox{0.9\hsize}{!}{\psfig{file=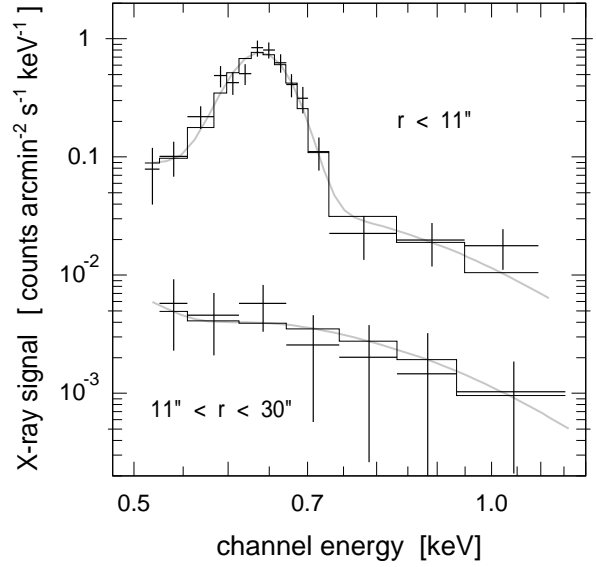,clip=}}}
\caption{X--ray spectra of Mars (top) and its X--ray halo (bottom).
Crosses with $1\sigma$ error bars show the observed spectra; the model
spectra, convolved with the detector response, are indicated by grey curves
(unbinned) and by histograms (binned as the observed spectra). The spectrum of
Mars itself is characterized by a single narrow emission line (this is most
likely the O--K$_{\alpha}$ fluorescence line; see the text for a discussion of
the apparent displacement of the line energy). At higher energies the presence
of an additional spectral component is indicated. The spectral shape of this
component can be well modeled by the same 0.2~keV thermal bremsstrahlung
emission which describes the spectrum of the X--ray halo.
}
\label{ms4spc}
\end{figure}

The X--ray spectra of Mars and its halo are shown in Fig.\,\ref{ms4spc}.
For the spectrum of Mars photons were extracted within $r<11''$ around its
center, for the X--ray halo photons within $11''<r<30''$ were used, and the
background was taken in both cases from an annulus around Mars with
$50''<r<100''$ (cf.\,Fig.\,\ref{ms4spa}).

\smallskip
The ACIS--I spectrum of Mars at
energies below $E\sim0.8\mbox{ keV}$ can be well described
($\chi_{\nu}^2=0.95$ for 10 degrees of freedom) by a single Gaussian emission
line at $E=0.65\pm0.01\mbox{ keV}$ with $\sigma=20\pm10\mbox{ eV}$ (i.e.,\ not
significantly broadened) and a flux of
$\left(6.3\pm0.8\right)\cdot10^{-5}\mbox{ ph cm}^{-2}\mbox{ s}^{-1}$.
Above energies of $\sim0.8\mbox{ keV}$, the presence of an additional
component is indicated. In the (instrumental) energy range
$E=0.5$\,--\,$1.1\mbox{ eV}$ (Fig.\,\ref{ms4spc}), the spectrum can be well
modeled ($\chi_{\nu}^2=0.89$ for 12 degrees of freedom) by a single Gaussian
emission line at $E=0.65\pm0.01\mbox{ keV}$, only instrumentally broadened
($\sigma\equiv0\mbox{ eV}$), with a flux of
$\left(5.4\pm0.9\right)\cdot10^{-5}\mbox{ ph cm}^{-2}\mbox{ s}^{-1}$,
superimposed on thermal bremsstrahlung with $kT$ fixed to 0.2~keV
(as for the halo; see below),
which contributes a flux of
$\left(1.5\pm0.4\right)\cdot10^{-5}\mbox{ ph cm}^{-2}\mbox{ s}^{-1}$, or
$\left(1.5\pm0.4\right)\cdot10^{-14}\mbox{ erg cm}^{-2}\mbox{ s}^{-1}$
in the energy range 0.5\,--\,1.2~keV.

\smallskip
The X--ray halo can be well characterized
by thermal bremsstrahlung emission with $kT=0.2\pm0.1\mbox{ keV}$
and a flux of
$\left(0.9\pm0.4\right)\cdot10^{-5}\mbox{ ph cm}^{-2}\mbox{ s}^{-1}$, or
$\left(0.9\pm0.4\right)\cdot10^{-14}\mbox{ erg cm}^{-2}\mbox{ s}^{-1}$
in the energy range 0.5\,--\,1.2~keV.
Further spectral analysis is limited by the poor photon statistics.

\subsection{Temporal variability}

\begin{figure}
\centerline{\resizebox{\hsize}{!}{\psfig{file=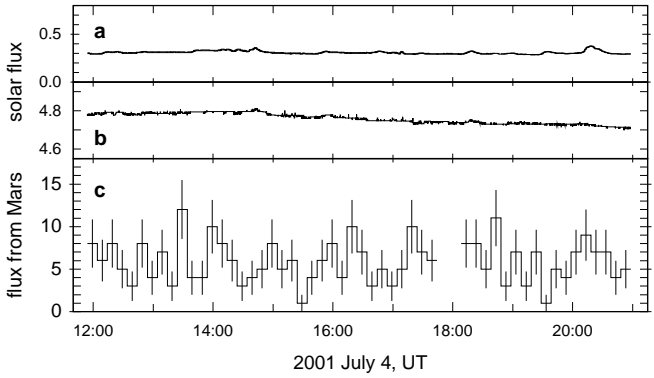,clip=}}}
\caption{Temporal behaviour of the soft X--ray flux from the Sun and
Mars. {\bf a)} 1\,--\,8~\AA\ (1.55\,--\,12.4~keV) solar flux in
$10^{-3}$ erg cm$^{-2}$ s$^{-1}$ at 1.0~AU, as measured with GOES--8
and GOES--10. {\bf b)} 1\,--\,500~\AA\ (0.025\,--\,12.4~keV)
solar flux in $10^{10}$ photons cm$^{-2}$ s$^{-1}$ scaled to 1.0~AU,
as measured with SOHO/SEM. The times in (a) and (b) were shifted by $+458$~s,
to take the light travel time delay between Sun\,$\to$\,Mars\,$\to$\,Chandra
and Sun\,$\to$\,SOHO/GOES into account.
{\bf c)} X--ray flux from Mars as observed with Chandra ACIS--I, in
counts/bin, shown with a bin size of 600~s, and derived by extracting
all photons below 1~keV from a circle of $11\farcs0$ radius centered at
Mars. The interruption at 18:00~UT is caused by Mars crossing the
gap between CCD I3 and I1. With $\sim0.1$ background events per time bin the
background is negligible (cf.\,Fig.\,\ref{ms4spa}).
}
\label{mlc}
\end{figure}

\begin{figure}
\centerline{\resizebox{0.9\hsize}{!}{\psfig{file=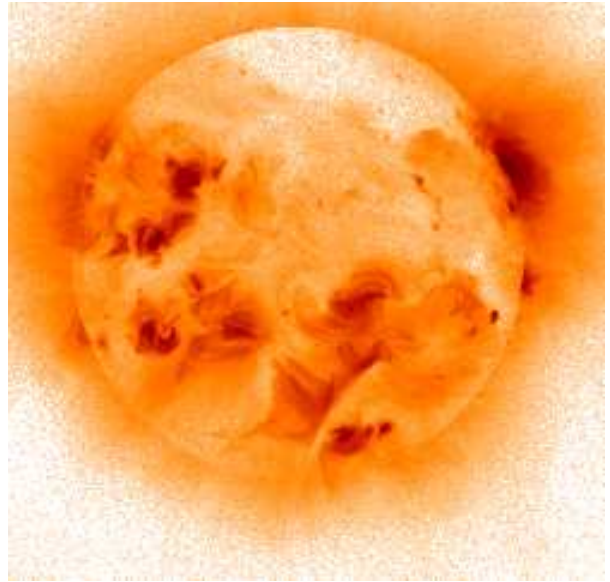,clip=}}}
\caption{Soft X--ray image of the Sun on 4 July 2001, 14:22~UT, obtained with
Yohkoh SXT in the energy range 0.25\,--\,4.0~keV
(http://mssly4.mssl.ucl.ac.uk/ydac/). It shows, within $8\degr$, the same
solar hemisphere which was irradiating Mars.
}
\label{yohkoh}
\end{figure}

The X--ray flux from Mars was fairly constant during the whole observation,
at $\left(9.3\pm0.6\right)\cdot10^{-3}\mbox{ counts s}^{-1}$ for $E<1\mbox{ keV}$
(Fig.\,\ref{mlc}c). According to the Kolmogorov--Smirnov test, the probability
that the observed count rates are statistical fluctuations around a constant
value is 30\%; the significance for intrinsic variability is only $1.3\,\sigma$.
The solar X--ray flux, monitored simultaneously with GOES--8 and GOES--10
(Fig.\,\ref{mlc}a) and SOHO/SEM (Fig.\,\ref{mlc}b) was also quite constant,
and unusually low for this phase in the solar cycle (Fig.\,\ref{goes02}).
These satellites observed within $8\degr$ the same solar hemisphere
which was irradiating Mars (Fig.\,\ref{yohkoh}).

\section{Modeling the X--ray appearance of Mars}

It was expected that fluorescent scattering of solar X--rays in the atmosphere
would be the dominant source of the X--ray radiation from Venus and Mars. In
order to get a reliable prediction about the X--ray properties of these
planets, a numerical model was developed for computing simulated images in the
individual fluorescence lines. This model was already successfully applied to
the Chandra observation of Venus \citep{02aap288}\<Dennerl..>. For Mars, it
was used in order to optimize the time of the observation (Sect.\,4.5). The
ingredients to the model are the composition and density structure of the Mars
atmosphere, the photoabsorption cross sections and fluorescence efficiencies
of the major atmospheric constituents, and the incident solar spectrum.

\subsection{Mars atmosphere}

\begin{figure}
\resizebox{\hsize}{!}{\psfig{file=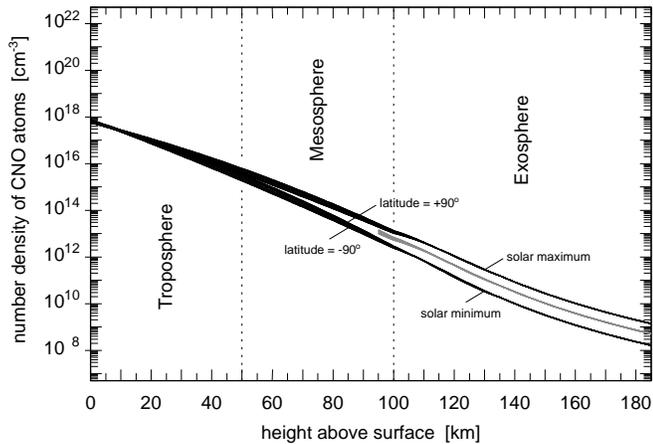,clip=}}
\caption{Number density $n_{\rm CNO}=n_{\rm C}+n_{\rm N}+n_{\rm O}$ of the sum
of C, N, and O atoms in the Mars model atmosphere as a function of the height
above the surface. Three sets of curves are shown: for solar minimum, solar
maximum, and the intermediate state (only above 95~km height, for better
clarity). Below 100~km, the density depends also on latitude.}
\label{atmo1}
\end{figure}

For the Mars atmosphere a simplified model was adopted, which describes the
total density $\rho$ in the form of analytical expressions for heights
0\,--\,100~km \citep{90bac002}\<Sehnal> and 100\,--\,1000~km
\citep{90bac001}\<Sehnal>. In order to get a smooth transition between both
regions, the density at 100\,--\,135~km heights was computed with both methods,
and the $\log\rho$ values were weighted according to their distance from 100
and 135~km. The analytical expressions are given for solar minimum, solar
maximum, and the intermediate state. For the simulation, the solar maximum
conditions were selected, motivated by the general behaviour of the soft solar
X--ray flux (Fig.\,\ref{goes02}). For simplicity it was assumed that the Mars
atmosphere is composed of C, N, and O only, neglecting the $\sim1.6\%$
contribution of other elements, mainly Ar,%
\footnote{Argon would produce a K$_{\alpha}$ fluorescence line at 2.96~keV
with a fluorescence yield $y_{\rm Ar}=0.118$ \citep{79jpc001}\<Krause>, which
exceeds that of nitrogen by a factor of 20. The photoabsorption cross section
of Ar--K$_{\alpha}$ is 15\% of that of N--K$_{\alpha}$, and there are about half
as many Ar atoms as N atoms, so that the X--ray albedo of Mars in the 
K$_{\alpha}$ fluorescence lines of N and Ar should be comparable.
The incident solar flux at 3.0~keV, however, is many orders of magnitude
lower than at 0.4~keV (Fig.\,\ref{crscflx}b), consistent with the
non--detection of any signal from Mars at 3.0~keV.
}
and the following composition was adopted: 64.9\% oxygen, 32.4\% carbon and
2.7\% nitrogen. As the main constituents, C and O, are contained in CO$_2$
(which accounts for more than 95\% of the Mars atmosphere), this composition
was assumed to be homogeneous throughout the atmosphere.

\subsection{Photoabsorption cross sections}

\begin{figure}
\resizebox{\hsize}{!}{\psfig{file=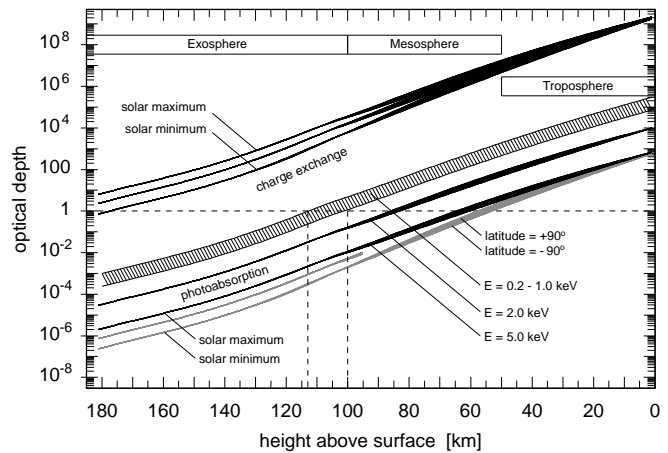,clip=}}
\caption{Optical depth $\tau=\tau_{\rm C}+\tau_{\rm N}+\tau_{\rm O}$
of the Mars model atmosphere with respect to charge exchange (above)
and photoabsorption (below), as seen from outside. The upper/lower
boundaries of the hatched area refer to energies just above/below the
C and O edges (cf.\ Fig.\,\ref{crscflx}a). For better clarity the
dependence of the photoabsorption on the solar cycle is only shown for
$E=5.0\mbox{ keV}$; the curves for the other energies refer to
solar maximum. The dashed horizontal line, at $\tau=1$, marks the
transition between the transparent ($\tau<1$) and opaque ($\tau>1$)
region. For charge exchange interactions a constant cross section of
$3\cdot10^{-15}\mbox{ cm}^2$ was assumed. Due to this high cross section,
$\tau=1$ is reached already at heights of 180~km and above, while for
photoabsorption at $E=0.2$\,--\,1.0~keV the atmosphere becomes opaque
between 113~km and 100~km, for solar maximum conditions. During solar
minimum, this transition occurs $\sim10\mbox{ km}$ deeper in the atmosphere.
}
\label{atmo2}
\end{figure}

The values for the photoabsorption cross sections were taken
from \cite{79apj362}\<Reilman and Manson>, supplemented by the
following K--edge energies (see \cite{02aap288}\<Dennerl..>
for a discussion of these energies):
$ E_{K_{\rm C}} = 296.1\mbox{ eV}, E_{K_{\rm N}} = 409.9\mbox{ eV},
  E_{K_{\rm O}} = 544.0\mbox{ eV}. $
From these values and the C, N, and O contributions listed above, the
effective photoabsorption cross section of the Mars atmosphere was computed
(Fig.\,\ref{crscflx}a). This, together with the atmospheric density structure,
yielded the optical depth of the Mars atmosphere, as seen from outside
(Fig.\,\ref{atmo2}).

\subsection{Solar radiation}

\begin{figure}
\resizebox{\hsize}{!}{\psfig{file=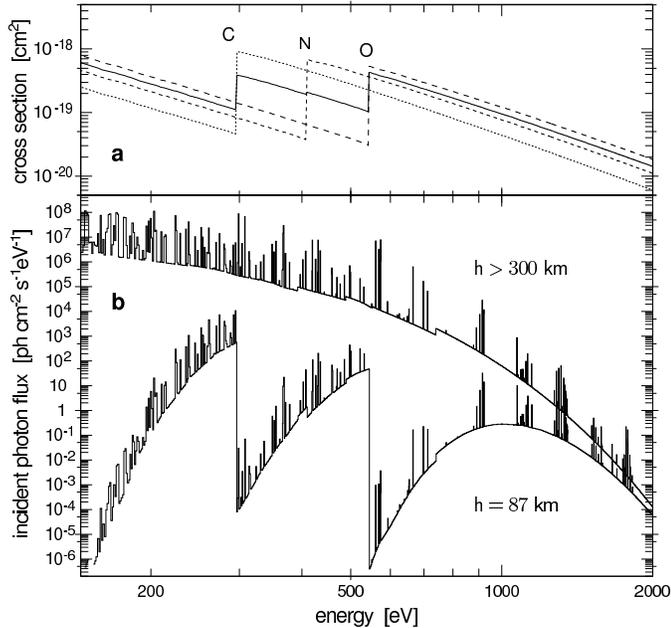,clip=}}
\caption{{\bf a)} Photoabsorption cross sections $\sigma_{\rm C}$,
$\sigma_{\rm N}$, $\sigma_{\rm O}$ for C, N, and O (dashed lines),
and $\sigma_{\rm CNO}$ for the chemical composition of the Mars
atmosphere (solid line).
{\bf b)} Incident solar X--ray photon flux on top of the Mars atmosphere
($h>300\mbox{ km}$) and at 87~km height. The spectrum is plotted in 1~eV bins.
At 87~km, it is considerably attenuated just above the K$_{\alpha}$ absorption
edges, recovering towards higher energies.
}
\label{crscflx}
\end{figure}

The solar spectra for 2001 July 4 were derived from SOLAR\,2000
\citep{00ast001}\<Tobiska..>.\footnote{available at {\tt
http://SpaceWx.com/}} To improve the coverage towards energies $E>100\mbox{
eV}$, synthetic spectra were computed with the model of
\cite{85aap287}\<Mewe..> and aligned with the SOLAR\,2000 spectra in
the range 50\,--\,500~eV, by adjusting the temperature and intensity. This
comparison yielded a fairly low average coronal temperature of only
$\sim80\mbox{ eV}$. The adopted solar spectrum, scaled to the heliocentric
distance of Mars, is shown in Fig.\,\ref{crscflx}\,b (upper curve), with a
bin size of 1~eV, which was used in order to preserve the spectral details.

\subsection{Model grid}

The high dynamic range in the optical depth of the Mars atmosphere requires a
model with high spatial resolution. Figure~\ref{atmo2} shows that the
atmosphere becomes optically thick for X--rays with $E<1\mbox{ keV}$ already
at heights above 100~km during solar maximum (and above 90~km during solar
minimum). This implies that most of the scattering takes place at heights where
the latitudinal dependence of the atmospheric density is negligible. Thus, the
volume elements need to be calculated only on a two dimensional grid (as in
the case of Venus). For the calculation a grid of cubic volume elements with a
side length of 1~km was used. The model atmosphere was traced from the surface
(at $r=3393.4\mbox{ km}$) to a height of 300~km. The simulated images were
synthesized with 20~km resolution perpendicular to the line of sight. Details
about the simulation itself can be found in \cite{02aap288}\<Dennerl..>.

\subsection{Planning the Mars observation}

\begin{figure}
\bigskip
\centerline{\resizebox{\hsize}{!}{\psfig{file=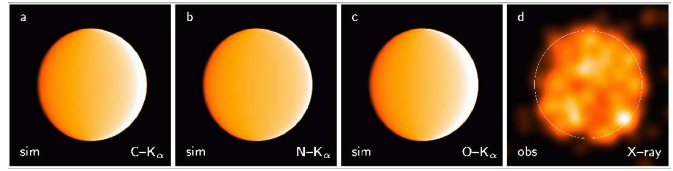,clip=}}}
\caption{{\bf a\,--\,c)} Simulated X--ray images of Mars at
C--K$_{\alpha}$, N--K$_{\alpha}$, and O--K$_{\alpha}$, for
a phase angle of $18\fdg2$.
The X--ray flux is coded in a linear scale,
extending from zero (black) to
{\bf a)} $5.0\cdot10^5\mbox{ ph cm}^{-2}\mbox{ s}^{-1}$,
{\bf b)} $1.1\cdot10^4\mbox{ ph cm}^{-2}\mbox{ s}^{-1}$, and
{\bf c)} $1.5\cdot10^5\mbox{ ph cm}^{-2}\mbox{ s}^{-1}$
(white). All images show some limb brightening,
especially at C--K$_{\alpha}$ and O--K$_{\alpha}$.
{\bf d)} Observed X--ray image, accumulated in the
energy range 0.4\,--\,0.7 and smoothed with a Gaussian filter with
$\sigma=1\farcs2$. The circle indicates the geometric size of Mars.
This image is dominated by O--K$_{\alpha}$ fluorescence photons.
Although the brightness fluctuations are mainly caused by photon
statistics and are not significant, there is evidence for limb brightening
on the right--hand side (cf.\,Fig.\ref{ms4spa}b).
}
\label{simsum}
\end{figure}

\begin{figure}
\resizebox{\hsize}{!}{\psfig{file=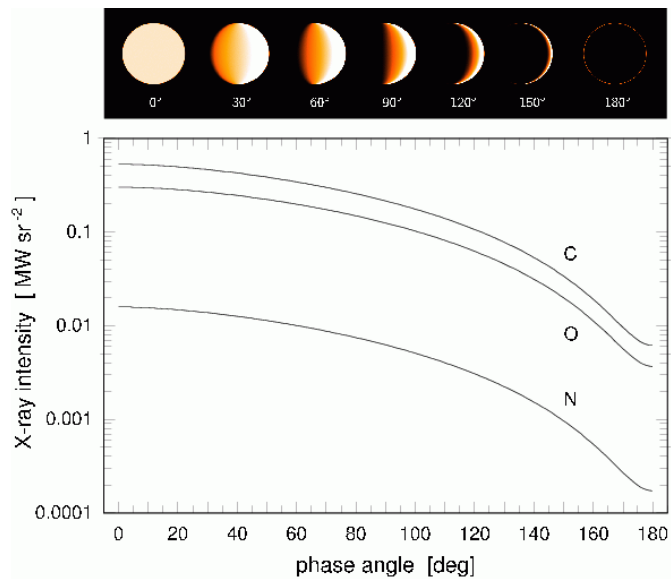,clip=}}
\caption{X--ray intensity of Mars as a function of phase angle, in the
fluorescence lines of C, N, and O, for the conditions on 4 July 2001. The
images at top, all displayed in the same intensity coding, illustrate the
appearence of Mars at O--K$_{\alpha}$ for selected phase angles.}
\label{mslum1}
\end{figure}

The simulation program was already used for optimizing the time of the Mars
observation. Although the closest approach of Mars to Earth, with a minimum
distance of 0.45~AU, occurred on 22 June 2002, the Chandra observation was
postponed by a few weeks. This decision was motivated by the fact that the
simulation indicated a practically uniform X--ray brightness across the whole
planet for this time (cf.\,Fig.\,\ref{mslum1}), while for phase angles of
$\sim15\degr$ and more, a diagnostically more valuable view was predicted,
with a characteristic brightening on the sunward limb
(Fig.\,\ref{simsum}a\,--\,c). The decision to postpone the Chandra observation
was supported by the favorable fact that Mars was still approaching the
perihelion of its orbit, so that its distance from Earth would increase only
slightly to 0.46~AU. Furthermore, the small loss of X--ray photons due to the
reduced solid angle would be almost compensated by the fact that Mars would
then be closer to the Sun and would intercept more solar radiation.

\section{Discussion}

\begin{figure}
\resizebox{\hsize}{!}{\psfig{file=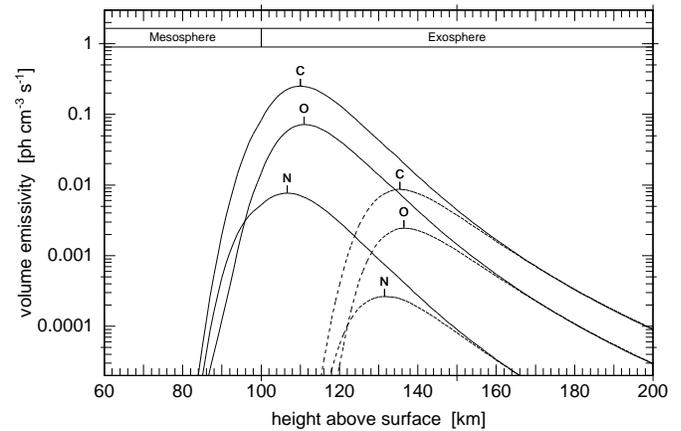,clip=}}
\caption{Volume emissivities of C, N, and O K$_{\alpha}$ fluorescent photons
at zenith angles of zero (subsolar, solid lines) and $90\degr$ (terminator,
dashed lines) for the incident solar spectrum of
Fig.\,\ref{crscflx}b. The height of maximum emissivity rises with
increasing solar zenith angles because of increased path length and
absorption along oblique solar incidence angles. In all cases maximum
emissivity occurs in the exosphere, where the optical depth depends
also on the solar cycle (Fig.\,\ref{atmo2}).
}
\label{volem}
\end{figure}

\subsection{Morphology}

The simulation shows (Fig.\,\ref{volem}) that the scattering of solar X--rays
takes place at heights above $\sim80\mbox{ km}$ and is most efficient between
110~km (along the subsolar direction) and 136~km (along the terminator). This
behaviour is similar to Venus, where the volume emissivity was found to peak
between 122~km and 135~km \citep{02aap288}\<Dennerl..>. The fact that the
volume emissivity for C is considerably higher than that of O is a direct
consequence of the unusually soft solar spectrum during the Mars observation
(cf.\,Fig.\,\ref{crscflx}b). During the Venus observation, the photon fluxes
from C and O were comparable.

\smallskip
Figures~\ref{simsum}a\,-\,c show the simulated images of Mars at the
K$_{\alpha}$ fluorescence lines of C, N, and O, for a phase angle of
$18\fdg2$. Although Mars was almost fully illuminated, there is
already some brightening on the more sunward limb evident, especially at C and
O, accompanied by a fading on the opposite limb. While a direct comparison
with the observed Mars image (Fig.\,\ref{simsum}d) suffers from low photon
statistics, a similar trend can be seen in the surface brightness profiles
(Fig.\,\ref{ms4spa}b). Thus, the expected limb brightening (Sect.\,4.5) was
actually observed. The reason for the limb brightening and the different
appearance of Mars in the three fluorescent lines is very similar to the case
of Venus, and a discussion can be found in \cite{02aap288}\<Dennerl..>.
The close match between the simulated and observed morphology is an argument
in favor of X--ray fluorescence as the dominant process responsible for the
X--ray radiation of Mars. With simulations based on charge exchange
interactions (Sect.\,5.5), \cite{01grl007}\<Holmstr"om> obtained a competely
different X--ray morphology.

\subsection{Spectrum, X--ray flux and luminosity}

The ACIS--I spectrum of Mars is dominated by a single narrow emission line.
Although this line appears at 0.65~keV, it is most likely the
O--K$_{\alpha}$ fluorescence line at 0.53~keV. This conclusion is motivated
by the fact that in the case of Venus a similar line was observed at 0.6~keV
with the same detector \cite[Fig.\,9 in][]{02prc002}\<Dennerl..>, which could
be uniquely identified to be at 0.53~keV by the additional LETG observation.
The apparent energy shift is most likely caused by optical loading, a
superposition of the charges released by 0.53~keV photons and optical photons,
during the 3.2~s exposure of each CCD frame.

\smallskip
The simulated images can be used to estimate the expected photon flux from
the whole visible side of Mars. For the three energies the following
values are obtained:
$f_{\rm C}=2.3\cdot10^{-4}$, 
$f_{\rm N}=5.0\cdot10^{-6}$, 
$f_{\rm O}=7.1\cdot10^{-5}\mbox{ ph cm}^{-2}\mbox{ s}^{-1}$.
While the C and N emission lines are outside the energy range
of ACIS--I, a direct comparison is possible for O--K$_{\alpha}$,
where a flux of
$\left(6.3\pm0.8\right)\cdot10^{-5}\mbox{ ph cm}^{-2}\mbox{ s}^{-1}$
was observed (Sect.\,3.2). This flux is reduced to
$\left(5.4\pm0.9\right)\cdot10^{-5}\mbox{ ph cm}^{-2}\mbox{ s}^{-1}$, if
an additional bremsstrahlung component is added. In view of all the general
uncertainties, these values are in good agreement with each other.

\smallskip
Conversion of the observed flux to the luminosity requires knowledge about
the angular distribution of the scattered photons. For this purpose,
X--ray intensities were determined from simulated Mars images,
computed for phase angles from $0\degr$ to $180\degr$ in steps of $1\degr$
(Fig.\,\ref{mslum1}). By spherically integrating these intensities for the
three energies over phase angle, the following luminosities are obtained from
the simulation: 2.9~MW for C, 0.1~MW for N, and 1.7~MW for O. The total X--ray
luminosity of Mars, 4.7~MW (or $\sim3.6\pm0.6\mbox{ MW}$ when adjusted to the
observed O--K$_{\alpha}$ flux), agrees well with the prediction of
\cite{01grl006}\<Cravens..>, who estimated a luminosity of 2.5~MW due to
X--ray fluorescence, with an uncertainty factor of about two. This is
another argument in favor of X--ray fluorescence.

\medskip
Compared to its optical flux, the X--ray flux of Mars is very low:
the visual magnitude $-2.1\mbox{ mag}$ corresponds to an optical flux
$f_{\rm opt} = 1.8\times10^{-4}\mbox{ erg cm}^{-2}\mbox{ s}^{-1}$.
Adopting a total X--ray flux
$f_{\rm x} = 9\times10^{-14}\mbox{ erg cm}^{-2}\mbox{ s}^{-1}$,
a ratio $f_{\rm x}/f_{\rm opt} = 5\times10^{-10}$ follows.
This is similar to the value $2\times10^{-10}$ observed for Venus
\citep{02aap288}\<Dennerl..>. In the case of X--ray fluorescence, the $f_{\rm
x}/f_{\rm opt}$ ratio of Mars is generally expected to exceed that of Venus,
because the optical albedo of Mars is lower than that of Venus, while their
X--ray albedos are comparable. Both ratios, however, are expected to vary with
time, in response to the temporarily variable solar X--ray flux
(cf.\,Fig.\,\ref{goes02}).

\subsection{The ROSAT observation of Mars in 1993}

\begin{figure}
\resizebox{\hsize}{!}{\psfig{file=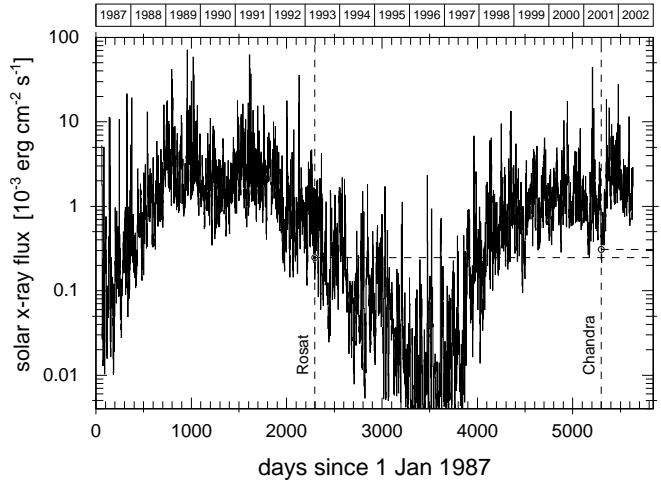,clip=}}
\caption{1\,--\,8~\AA\ (1.55\,--\,12.4~keV) solar X--ray flux at 1.0~AU,
measured with GOES--7 (before March 1995) and GOES--8 (afterwards).
During the ROSAT and Chandra observations (indicated by dashed vertical
lines), the solar X--ray flux was similar.
}
\label{goes02}
\end{figure}

\begin{figure*}[ht]
\hbox to \hsize{\vbox{
\vbox{\vsize=0.30\hsize
\psfig{file=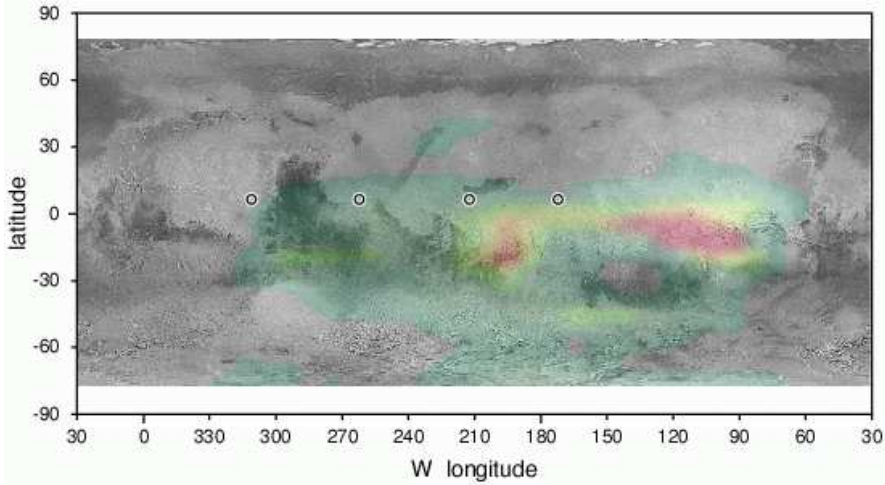,height=1.2\vsize}
}}\hspace*{6mm}\raisebox{-1mm}{\vbox{\hsize=0.3\hsize
\smallskip
\caption{Mars during the Chandra observation. {\bf Grey:} Viking map of Mars,
courtesy NASA/JPL/Caltech (http://maps.jpl.nasa.gov/). For a better
comparison with the simulated and observed images, this map was shifted by
150$^{\circ}$ in longitude.
The circles indicate the central positions of the images in
Figs.\,\ref{msim} and \ref{mobs}.
{\bf Color:} Dust optical depth, derived with the Thermal Emission
Spectrometer on Mars Global Surveyor from 12 orbits on 4 July 2001
(adopted from Smith et al.\ 2002). Green, yellow, and red colours mark
dust optical depths of 1.0, 1.5, and $\ge2.0$. In all coloured areas the
atmosphere was very dusty and in a typical dust storm condition.
}
\label{marsadd}
}
}}
\end{figure*}

Mars was observed with the ROSAT Position Sensitive Proportional Counter
(PSPC) from 10\,--\,13 April 1993 on three occasions, for 1\,294~s, 2\,124~s,
and 1\,099~s, respectively. As the pointing direction of the satellite was kept
fixed during these observations, Mars was located at different positions in the
$2\degr$ PSPC field of view (FOV). During the first and third exposure,
Mars was partially obscured by a radial strut of the PSPC support structure,
and was furthermore placed so much in the outer parts of the FOV, where the
point spread function
was severely degraded, that only the second observation is suited for a
sensitive search for any X--ray emission. During this observation, Mars was at
a heliocentric distance of 1.67~AU. The constraint to observe it at an
elongation close to $90\degr$ implied a fairly large geocentric distance from
Earth, 1.32~AU, so that Mars appeared as a disk with a diameter of only
$7\farcs1$, seen at a phase angle of $37\fdg0$. Mars was not detected in this
observation. From the second PSPC exposure, a $3\sigma$ upper limit of
$4\cdot10^{-3}\mbox{ counts s}^{-1}$ can be derived in the energy range
0.1\,--\,0.9~keV, for a circle around the nominal position of Mars with a
radius of $1'$.

How does this non--detection with ROSAT compare with the information
which is now available on the X--ray properties of Mars\,?
According to the SOLAR\,2000 data for 13 April 1993, the solar X--ray flux at
1~AU was about 30\% fainter than during the Chandra observation
(cf.\,Fig.\,\ref{goes02}), but showed a similar spectral shape. Taking
also the larger heliocentric distance into account, Mars received about half
of the solar flux. The simulation then yields
the following number of expected counts for the second PSPC observation:
$5\mbox{ counts}$ at C, $0.003\mbox{ counts}$ at N, and $0.3\mbox{ counts}$
at O. The total number of expected counts, $\sim5$, is somewhat lower than the
local background in the detect cell ($\sim8$). This suggests that the X--ray
signal of Mars was just below the sensitivity limit of the ROSAT observation.

\subsection{Temporal variability and the dust storm of 2001}

Scattering of solar X--rays on very small dust particles was one of the early
suggestions for explaining the X--ray emission from comets.
\cite{96ass009}\<Wickramasinghe and Hoyle> noted that X--rays can be
efficiently scattered by dust particles, if their size is comparable to the
X--ray wavelength. Such attogram dust particles ($\sim10^{-18}\mbox{ g}$)
would be difficult to detect by other means. It might be possible that such
particles are present in the upper Mars atmosphere, in particular during
episodes of global dust storms.


\smallskip
Incidentally, on June 26 a local dust storm on Mars originated and expanded
quickly, developing into a planet--encircling dust storm by July 11
\citep{02ica015}\<Smith..>. Such dust storms have been observed on roughly
one--third of the perihelion passages during the last decades, but never so
early in the Martian year. On July 4, this very vigorous storm had covered
roughly one hemisphere (Fig.\,\ref{marsadd}). This hemisphere happened to be
visible at the beginning of the Chandra observation. By the end of the
observation, which covered one third of a Mars rotation, this hemisphere had
mainly rotated away from our view (Figs.\,\ref{marsadd}\,--\,\ref{mobs}).
Thus, a comparison of the Chandra data from both regions should reveal any
influence of the dust storm on the X--ray flux.

\smallskip
There is, however, no change in the mean X--ray flux between the first and
second half of the observation, where 150 and 157 photons were detected,
respectively. This implies that, if attodust particles are present in the
upper Mars atmosphere, the dust storm did not lead to a local increase in
their density, high enough to modify the observed X--ray flux significantly.
No statement, however, can be made about the situation below $\sim80\mbox{ km}$,
as the solar X--rays do not reach these atmospheric layers (Fig.\,\ref{volem}).
While the general presence of some attodust in the upper atmosphere cannot be
ruled out
by the Chandra observation, the fact that the ACIS--I spectrum of Mars is
dominated by a single emission line (Fig.\,\ref{ms4spc}) shows that any
contribution of such particles to the X--ray flux from Mars must be small
compared to fluorescence, even in the process of a developing global dust
storm.


\begin{figure}
\psfig{file=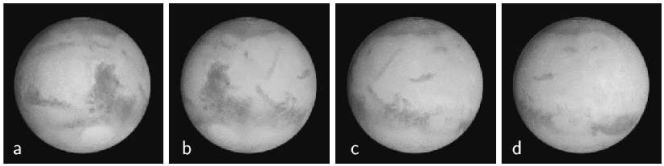,width=\hsize}
\caption{Simulated ``clear" views of Mars, obtained from
http://space.jpl.nasa.gov/.
In order to facilitate comparison with the map
(Fig.\,\ref{marsadd}), they were arranged
with decreasing central meridian.
{\bf a)} 04 July 2001, 21:40 UT, $\mbox{LCM}=311\degr$
{\bf b)} 01 July 2001, 16:30 UT, $\mbox{LCM}=262\degr$
{\bf c)} 04 July 2001, 14:55 UT, $\mbox{LCM}=212\degr$
{\bf d)} 04 July 2001, 12:10 UT, $\mbox{LCM}=172\degr$.
}
\label{msim}

  \end{figure}
  \begin{figure}[ht]

%
\psfig{file=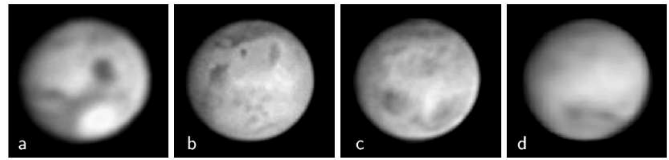,width=\hsize}
\caption{Optical images of Mars, taken during the Chandra observation
(except b, which was taken three days earlier), and arranged as in
Fig.\,\ref{msim}.
{\bf a)} 04 July 2001, 21:38 UT, $\mbox{LCM}=311\degr$
courtesy B.\,Flach--Wilken, Germany
{\bf b)} 01 July 2001, 16:30 UT, $\mbox{LCM}=262\degr$,
courtesy T.\,Wei\,Leong, Singapore
{\bf c)} 04 July 2001, 14:55 UT, $\mbox{LCM}=212\degr$,
courtesy Y.\,Morita, Japan,
CMO Archives of the OAA Mars Section
{\bf d)} 04 July 2001, 12:10 UT, $\mbox{LCM}=172\degr$,
courtesy G.\,Garradd, Australia.
Note how significantly the surface markings were changed by the dust storm,
particularly at longitudes $\ell_{\rm w}<270\degr$ (b\,--\,d), while the
hemisphere at $\ell_{\rm w}>270\degr$ (a) was much less affected.
}
\label{mobs}
\end{figure}

\subsection{The X--ray halo}

Although the significance of a soft X--ray halo around Mars is only
$\sim4\sigma$, its spectrum is clearly different from that of Mars itself
(Fig.\,\ref{ms4spc}), ruling out the possibility that the halo is an
instrumental artefact, related to the point spread function of the X--ray
telescope. It can also be ruled out that the halo is caused by the vignetting
of the telescope, because the $11''<r<30''$ halo contains $2.1\pm0.3$ times
more photons than the same area in the $50''<r<100''$ background region, while
vignetting would affect the number of photons by less than 5\% at energies
below 1.5~keV. Furthermore, the halo cannot be an artefact of exposure
variations introduced by removing the point sources, because no gradient in
the surface brightness is observed at $E=1.5$\,--\,$10.0\mbox{ keV}$
(Fig.\,\ref{ms4spa}b). Therefore, the following discussion assumes that the
X--ray halo is real.

\smallskip
While there is a lot of evidence that the X--rays from Mars are predominantly
caused by fluorescent scattering of solar X--rays in its upper atmosphere,
there is the possibility of an additional source of X--ray emission. When
highly ionized heavy ions in the solar wind encounter atoms in the exosphere
of Mars, they become discharged and may emit X--rays. This is the process
which was found to be responsible for the X--ray emission of comets
\citep{97grl001,01sci008}\<Cravens>\<Lisse..>. Its consequences for the X--ray
emission of Mars were already investigated by several authors.
\cite{00asr006}\<Cravens> predicted an X--ray luminosity of
$\sim0.01\mbox{ MW}$.
\cite{00ica016}\<Krasnopolsky> estimated an X--ray emission of
$\sim4\times10^{22}\mbox{ ph s}^{-1}$.
Adopting an average photon energy of 200~eV \cite[e.g.][]{97grl001}\<Cravens>,
this corresponds to an X--ray luminosity of 1.3~MW.
\cite{01grl007}\<Holmstr\"om..> computed a total X--ray luminosity of Mars due
to charge exchange (within 10 Mars radii) of 1.5~MW at solar maximum, and
2.4~MW at solar minimum. 

\smallskip
For the X--ray halo observed within 3 Mars radii, excluding Mars itself,
the Chandra observation yields a flux of
$\left(0.9\pm0.4\right)\times10^{-14}\mbox{ erg cm}^{-2}\mbox{ s}^{-1}$
in the energy range $E=0.5$\,--\,$1.2\mbox{ keV}$ (Sect.\,3.2).
Assuming isotropic emission, this flux corresponds to a luminosity of
$0.5\pm0.2\mbox{ MW}$. This value agrees well with the predictions of
\cite{00ica016}\<Krasnopolsky> and \cite{01grl007}\<Holmstr\"om..>, in
particular when the spectral shape is extrapolated to lower energies.%
\footnote{It has, however, be kept in mind that there may be an additional
uncertainty in these values, because the O--K$_{\alpha}$ fluorescence line
was found to be displaced by $\sim120\mbox{ eV}$ (Sect.\,5.2).}

\smallskip
In addition to the luminosity, there is another argument in favor of the idea
that the X--ray halo may be the signature of charge exchange. Although
this process produces a spectrum consisting of many narrow emission lines, the
overall properties can be approximated by 0.2~keV thermal bremsstrahlung
emission \citep{98pss001}, and the spectrum of the X--ray halo agrees very
well with such a model. Also the spectrum of Mars itself shows evidence for an
emission component with this spectral shape (Fig.\,\ref{ms4spc}). The Chandra
data, however, indicate that the surface brightness of this component in the
spectrum of Mars is by one order of magnitude higher than that in the halo,
averaged from one to three Mars radii. This is different from the result of
computer simulations by \cite{01grl007}\<Holmstr\"om..>, where the surface
brightness in front of Mars is lower than in the halo. In these simulations an
empirical model of the proton flow near Mars was used, where the proton flux
decreases strongly at the ``magnetopause'', at $\sim680\mbox{ km}$ height.
The fact that the surface brightness at the center was observed to be higher
than expected could be an indication that the dilution of the heavy ion flux
near the ``magnetopause'' might be less pronounced than assumed in the model.
It has to be stressed, however, that the observational evidence for any
emission component in addition to the X--ray fluorescence is near the
sensitivity limit of the observation and that any statement about
observational properties may be subject to considerable uncertainties.

\section{Summary and conclusions}

The Chandra observation clearly shows that Mars is an X--ray source. The
luminosity, the X--ray spectrum, the morphology and the time variability are
all consistent with fluorescent scattering of solar X--rays on oxygen atoms in
the Mars atmosphere at heights above $\sim80\mbox{ km}$ as the main process
for the observed radiation. No evidence for dust--related X--ray emission was
found, despite the onset of a global dust storm, which had covered roughly one
hemisphere at the time of the observation. Differential measurements between
the hemisphere affected by the dust storm and the quiet hemisphere showed no
significant difference in the X--ray flux. There is, however, some evidence
for an additional source of X--ray emission, indicated by a faint X--ray halo
which can be traced to about three Mars radii, and by an additional component
in the X--ray spectrum of Mars, which has a similar spectral shape as the halo.
Within the available limited statistics, the spectrum of this component can be
characterized by 0.2~keV thermal bremsstrahlung emission. The spectral shape
and the luminosity are indicative of charge exchange interactions between highly
charged heavy ions in the solar wind and exospheric hydrogen and oxygen around
Mars. The significance of the halo, however, is only $4\sigma$, and additional
observations will be needed for further studies. Such observations would also
provide additional information about the temporal properties of the exosphere
of Mars, in particular with respect to the solar cycle.

\begin{acknowledgements}
It is a great pleasure to thank S.\,Wolk for his support in planning this
observation, B.\,Aschenbach, V.\,Burwitz, J.\,Englhauser and C.\,Lisse for
stimulating discussions, and G.\,Garradd, Y.\,Morita, T.\,Wei\,Leong and
B.\,Flach--Wilken for providing the optical images. SOLAR\,2000 Research Grade
v1.15 historical irradiances are provided courtesy of W.~Kent Tobiska and
SpaceWx.com. These historical irradiances have been developed with funding
from the NASA UARS, TIMED, and SOHO missions. The SOHO CELIAS/SEM data were
provided by the USC Space Sciences Center. SOHO is a joint European Space
Agency, United States National Aeronautics and Space Administration mission.
The Yohkoh image was obtained from the Yohkoh Data Archive Centre (YDAC).
Yohkoh is a mission of the Japanese Institute for Space and Astronautical
Science. 
\end{acknowledgements}


\end{document}